\documentclass[11pt]{article} 
\usepackage{amsfonts,amsmath,amsxtra}
\usepackage{latexsym}
\usepackage{amssymb}

\def\hybrid{\topmargin 0pt      \oddsidemargin 0pt
        \headheight 0pt \headsep 0pt
        \textwidth 16.5cm
        \textheight 23cm
        \marginparwidth 0.0in
        \parskip 5pt plus 1pt   \jot = 1.5ex}
\catcode`\@=11
\def\marginnote#1{}
\newcount\hour
\newcount\minute
\newtoks\amorpm
\hour=\time\divide\hour by60 \minute=\time{\multiply\hour by60
\global\advance\minute by-\hour}
\edef\standardtime{{\ifnum\hour<12 \global\amorpm={am}%
        \else\global\amorpm={pm}\advance\hour by-12 \fi
        \ifnum\hour=0 \hour=12 \fi
      \number\hour:\ifnum\minute<10 0\fi\number\minute\the\amorpm}}
\edef\militarytime{\number\hour:\ifnum\minute<10 0\fi\number\minute}

\def\draftlabel#1{{\@bsphack\if@filesw {\let\thepage\relax
   \xdef\@gtempa{\write\@auxout{\string
      \newlabel{#1}{{\@currentlabel}{\thepage}}}}}\@gtempa
   \if@nobreak \ifvmode\nobreak\fi\fi\fi\@esphack}
        \gdef\@eqnlabel{#1}}
\def\@eqnlabel{}
\def\@vacuum{}
\def\draftmarginnote#1{\marginpar{\raggedright\scriptsize\tt#1}}

\def\draft{\oddsidemargin -0.1truein
        \def\@oddfoot{\sl preliminary draft \hfil
        \rm\thepage\hfil\sl\today\quad\militarytime}
        \let\@evenfoot\@oddfoot \overfullrule 3pt
        \let\label=\draftlabel
        \let\marginnote=\draftmarginnote
\def\@eqnnum{{\rm (\theequation)}
\rlap{\kern\marginparsep\tt\@eqnlabel}%
\global\let\@eqnlabel\@vacuum}  }

\newfont{\Bbbb}{msbm7 scaled 1\@ptsize00}
\newcommand{\zs}{\raise-1pt\hbox{$\mbox{\Bbbb Z}$}}

scaled 1\@ptsize00

\font\sevenmsa=msam6 
\newfam\msafam
\textfont\msafam=\sevenmsa
\def\hexnumber@#1{\ifnum#1<10 \number#1\else
\ifnum#1=10 A\else\ifnum#1=11 B\else\ifnum#1=12 C\else \ifnum#1=13
D\else\ifnum#1=14 E\else\ifnum#1=15 F\fi\fi\fi\fi\fi\fi\fi}
\def\msa@{\hexnumber@\msafam}
\def\llcorner{\delimiter"4\msa@78\msa@78 }
\def\lrcorner{\delimiter"5\msa@79\msa@79 }
\mathchardef\blacktriangleright="3\msa@49
\mathchardef\blacktriangleleft="3\msa@4A \font\tenmsb=msbm10 scaled
1\@ptsize00
\newfam\msbfam
\textfont\msbfam=\tenmsb \scriptfont\msbfam=\tenmsb

\newdimen\Squaresize \Squaresize=14pt
\newdimen\Thickness \Thickness=0.5pt

\def\Square#1{\hbox{\vrule width \Thickness
   \vbox to \Squaresize{\hrule height \Thickness\vss
      \hbox to \Squaresize{\hss#1\hss}
   \vss\hrule height\Thickness}
\unskip\vrule width \Thickness} \kern-\Thickness}

\def\Vsquare#1{\vbox{\Square{$#1$}}\kern-\Thickness}

\def\numberbysection{\@addtoreset{equation}{section}
        \def\theequation{\thesection.\arabic{equation}}}
\numberbysection

\renewcommand{\theequation}{\thesection.\arabic{equation}}
\def\titlepage{\@restonecolfalse\if@twocolumn\@restonecoltrue\onecolumn
     \else \newpage \fi \thispagestyle{empty}\c@page\z@
        \def\thefootnote{\fnsymbol{footnote}} }

\def\endtitlepage{\if@restonecol\twocolumn \else  \fi
        \def\thefootnote{\arabic{footnote}}
        \setcounter{footnote}{0}}  
\relax

\hybrid
\makeatletter
\newdimen\normalarrayskip            
\newdimen\minarrayskip               
\normalarrayskip\baselineskip \minarrayskip\jot
\newif\ifold             \oldtrue            \def\new{\oldfalse}
\def\arraymode{\ifold\relax\else\displaystyle\fi}
\def\eqnumphantom{\phantom{(\theequation)}} 
\def\@arrayskip{\ifold\baselineskip\z@\lineskip\z@
     \else
     \baselineskip\minarrayskip\lineskip1\baselineskip\fi}

\def\@arrayclassz{\ifcase \@lastchclass \@acolampacol \or
\@ampacol \or \or \or \@addamp \or
   \@acolampacol \or \@firstampfalse \@acol \fi
\edef\@preamble{\@preamble
  \ifcase \@chnum
     \hfil$\relax\arraymode\@sharp$\hfil
     \or $\relax\arraymode\@sharp$\hfil
     \or \hfil$\relax\arraymode\@sharp$\fi}}

\def\@array[#1]#2{\setbox\@arstrutbox=\hbox{\vrule
     height\arraystretch \ht\strutbox
     depth\arraystretch \dp\strutbox
width\z@}\@mkpream{#2}\edef\@preamble{\halign \noexpand\@halignto
\bgroup \tabskip\z@ \@arstrut \@preamble \tabskip\z@ \cr}%
\let\@startpbox\@@startpbox \let\@endpbox\@@endpbox
  \if #1t\vtop \else \if#1b\vbox \else \vcenter \fi\fi
  \bgroup \let\par\relax
  \let\@sharp##\let\protect\relax
  \@arrayskip\@preamble}
\def\eqnarray{\stepcounter{equation}%
              \let\@currentlabel=\theequation
              \global\@eqnswtrue
              \global\@eqcnt\z@
              \tabskip\@centering              
              \let\\=\@eqncr
              $$%
            \halign to \displaywidth  \bgroup
             \eqnumphantom \@eqnsel
      \hskip\@centering                               
    $\displaystyle  \tabskip\z@ {##}$%
    &\global\@eqcnt\@ne \hskip 2\arraycolsep
         $ \displaystyle  \arraymode{##}$\hfil
    &\global\@eqcnt\tw@ \hskip 2\arraycolsep
         $\displaystyle\tabskip\z@{##}$\hfil
         \tabskip\@centering
    &{##}\tabskip\z@\cr}
\makeatother

\newcommand{\CC}{{\mathbb{C}}}

\def\IC{\mathbb{C}}

\def\IP{\mathbb{P}}

\def\IR{\mathbb{R}}
\def\IZ{\mathbb{Z}}
\def\CA {\mathcal{A}}

\def\CC {\mathcal{C}}

\def\CF {\mathcal{F}}
\def\CG {\mathcal{G}}

\def\CL {\mathcal{L}}
\def\CM {\mathcal{M}}
\def\CN {\mathcal{N}}
\def\CO {\mathcal{O}}
\def\CP {\mathcal{P}}

\def\CS {\mathcal{S}}

\def\CV {\mathcal{V}}

\def\CZ {\mathcal{Z}}

\def\a {{\alpha}}

\def\pr {\partial}
\def\apr {\overline {\partial }}

\def\jb{\bar{j}}

\def\zb {\bar{z}}

\def\Tr{{\rm Tr}}

\def\frak{\mathfrak}
\def\Fg{{\frak g}}

\newcommand\bqa{\begin{eqnarray}}
\newcommand\eqa{\end{eqnarray}}
\def\be{\begin{eqnarray}\new\begin{array}{cc}}
\def\ee{\end{array}\end{eqnarray}}

\def\beq{\begin{equation}}
\def\eeq{\end{equation}}
\def\bse{\begin{subequations}}                
\def\ese{\end{subequations}}
\def\bp{\begin{pmatrix}}
\def\ep{\end{pmatrix}}

\def\i{\imath}

\def\stack#1#2{\raise0.7pt\hbox{$\mathrel{\mathop{#2}\limits^{#1}}$}}
\def\tr{\triangleright}
\def\tl{\triangleleft}
\def\sem{\mathsurround=0pt \raise1pt
\hbox{$\scriptscriptstyle>\!\!$}\:\!\!\tl}
\def\mes{\mathsurround=0pt \tr\!\:\!\raise0.8pt
\hbox{$\scriptscriptstyle\!\!<$}\,}
\def\]{\mathsurround=0pt ]\raise-2pt\hbox{$_\ast$}}

\def\<{\langle}
\def\>{\rangle}

\def\frak{\mathfrak}

\def\CO{{\cal O}}

\def\CZ{{\cal Z}}

\def\we{\raise-1pt\hbox{$\,\stackrel{\wedge}{,}\,$}}
\def\tr{{\rm tr}\,}
\def\Tr{{\rm Tr}\,}
\def\pr {\partial}

\newcounter{pac}[section]

\newcounter{pacc}[subsection]

0

0

\title{\bf  On topological field theory  representation \\ of \\
 higher analogs of classical special functions
\footnote{Extended version of a talk given by the
first author at {\it Quantum field theory
and representation theory}, October, 2010, Moscow, Russia.}}

\begin{document}
\author{Anton A. Gerasimov and  Dimitri R. Lebedev}
\date{}

\maketitle

\renewcommand{\abstractname}{}

\begin{abstract}

\noindent {\bf Abstract}.
Looking for a quantum field theory model of 
Archimedean algebraic geometry a class of infinite-dimensional
integral representations of  classical special functions 
was introduced. Precisely the special functions
 such as Whittaker functions and $\Gamma$-function
were identified with correlation functions in
topological field theories on a two-dimensional disk.
Mirror symmetry of the underlying topological field theory 
leads to a dual finite-dimensional
integral representations reproducing classical integral representations for
the corresponding special functions. The 
mirror symmetry interchanging infinite- and finite-dimensional integral
representations provides  an incarnation of the 
local Archimedean Langlands duality on the level of classical special functions.

In this note we provide some directions to higher-dimensional generalizations of
our previous results.  In the first part we consider topological field theory
representations of multiple local $L$-factors introduced by
Kurokawa and expressed through multiple Barnes's  $\Gamma$-functions.   
In the second part we are dealing with 
generalizations based on consideration of
 topological Yang-Mills theories  on non-compact four-dimensional manifolds.
Presumably, in analogy with the mirror duality in two-dimensions, 
$S$-dual  description
should be instrumental for deriving integral representations
for a particular class of quantum field theory correlation functions
and  thus  providing  a new 
interesting class of  special functions supplied with canonical integral
representations.

\end{abstract}
\vspace{5 mm}

\maketitle
\renewcommand{\abstractname}{}

\section*{Introduction}

In \cite{GLO2}, \cite{GLO3}, \cite{GLO4}, \cite{GLO5}, \cite{G}
 a topological field theory framework for a
description of the Archimedean algebraic geometry was proposed.
As a first step \cite{GLO2}, \cite{GLO3} local Archimedean $L$-factors
were interpreted as  correlation functions in
two-dimensional equivariant topological field theories
on a disk. It was demonstrated that
the local Archimedean Langlands correspondence
between various constructions of $L$-factors
 (see e.g. \cite{B}, \cite{L})  
is realized as  mirror symmetry on the level of  underlying
topological field theories \cite{GLO3}. These results were generalized to a class of
Whittaker functions in \cite{GLO4}.  Presumably this picture holds in
full generality and provides a realization of the Archimedean
Langlands duality for a generic Whittaker function. 
Moreover, one can expect
that the approach of \cite{GLO2}, \cite{GLO3}, \cite{GLO4},
\cite{GLO5}, \cite{G} will lead to an interpretation of all  basic
constructions of Archimedean algebraic geometry in terms of
two-dimensional topological field theories thus 
providing a clue to a natural formulation of the geometry 
over Archimedean fields.

Given the  proposed connection of the Archimedean geometry with 
two-dimensional topological field theories it is natural to ask what
is a special role of two dimensions in these considerations     
and are  there any signs of 
possible generalizations  of \cite{GLO2}, \cite{GLO3}, \cite{GLO4}
to other dimensions. The
case of zero dimension was considered  in \cite{GL1}.
This note is a very preliminary discussion of a large project of
higher dimensional generalizations  of \cite{GLO2}, \cite{GLO3}, \cite{GLO4}.
In the first part we propose a description of higher analogs of 
local Archimedean $L$-factors introduced
by Kurokawa  \cite{Ku1}, \cite{Ku2} (see also \cite{Ma}) in terms of equivariant
topological  field theories with quadratic actions. The Kurokawa 
$L$-factors shall  be considered as basic 
building blocks of higher-dimensional generalizations
of the Mellin-Barnes representations \cite{KL}. We also speculate on a dual
description generalizing  type $B$  description of the standard local
Archimedean $L$-factors \cite{GLO3}. Eventually this should lead to a
 higher-dimensional generalization of the local Archimedean
Langlands correspondence.

In the second part of this note we pursue another (but related) 
direction of  generalization
of \cite{GLO4} relying on the 
four-dimensional topological field theories obtained
by twisting of  $\CN=2$ SUSY Yang-Mills theory on non-compact
four-dimensional manifolds. These theories 
have $S$-dual descriptions in terms of
 theories  of abelain gauge fields interacting with
monopoles.  There exists  a class of  correlation functions 
in twisted $\CN=2$ SUSY Yang-Mills theories on non-compact
four-dimensional manifolds 
computable by direct counting gauge theory instantons
\cite{LNS}, \cite{N}, \cite{NO}, \cite{NY}. 
This class of  correlation functions 
shall be considered as  a close  analog of 
the class of correlation functions in type $A$
topological sigma models considered in  \cite{GLO4}.
Thus this  class of correlation
functions in four-dimensional topological field theories 
should also have natural integral representations similar
to those derived in \cite{GLO4} for two-dimensional topological sigma
models. In two-dimensional case the integral representations 
of correlation functions in  topological sigma models 
provide a direct relation with the mirror dual formulations in terms of
topological Landau-Ginzburg theories. One expects that in
four-dimensions  the integral representations of
the instanton counting functions provide  direct links with 
the dual monopole descriptions of the gauge theories  (captured
effectively by the Seiberg-Witten prepotential). This should 
also provide  integral representations of  new special
functions generalizing Whittaker functions (see \cite{BE} for
 related considerations). In this note we discuss what
can be considered as a proper set-up for a verification of these
hopes. We consider only the case of 
 non-compact four-manifolds with an action
of $S^1$ isometry leaving more general cases e.g. allowing  actions of $S^1\times
S^1$ for future publication \cite{GL2}. Meanwhile, as a simple
exercise,  we explicitly construct an  integral representation
of the Mellin-Barnes type for a limit of
equivariant instanton counting function (first introduced in
\cite{LNS}) describing vortex in two-dimensional models (see e.g. \cite{Shad}).

Let us make a short comment on  (a small part of)  modern literature on
topological / supersymmetric quantum field theories relevant to the
subject of this note. One of the key points of the constructions
of \cite{GLO2}, \cite{GLO3}, \cite{GLO4} was the use the equivariant
setting with respect to a group of global symmetries including
space-time rotations.  Although a relation between $S^1$-equivariance
and coupling of two-dimensional topological sigma models 
with topological gravity was  well-known,  
the effective  use of the $S^1$-equivaraint extensions
of topological sigma  model
was initiated by Kontsevich \cite{K} and further advanced in \cite{Gi1},
\cite{Gi2}, \cite{Gi3}. The four-dimensional analog of this approach 
was considered  in \cite{LNS}  in the context of
instanton counting on $\IR^4$ and was convincingly related in \cite{N}, \cite{NO}, 
\cite{NY} to  the  Seiberg-Witten solution \cite{SW}
of $\CN=2$ SUSY Yang-Mills theory.
An identification of a correlation function on a disk in
a class of two-dimensional topological field theories
with solutions of quantum integrable systems was proposed in
\cite{GS1} using previous findings in \cite{MNS}
and was argued in \cite{GS2} to be a general phenomena. Many examples, including
four-dimensional cases relevant to considerations of this note were
considered in \cite{NS} (see also \cite{NW}).
For a detailed discussion of supersymmetric/ topological
quantum field theories on non-compact manifolds  see \cite{GW}.
In the  remarkable paper \cite{AGT} a relation between correlation
functions of four-dimensional SUSY gauge theories and correlation
functions in two-dimensional models was proposed. This in particular 
implies that the class of special functions we are looking for  shall include
building blocks of correlation functions in two-dimensional theories (such as
conformal Toda theories). Note also that
counting  of BPS states in SUSY quantum field theories \cite{R1},
\cite{KMMR}, \cite{R2} leads to  generalized Mellin-Barnes type
integral representations in terms of
combinations of double $\Gamma$-functions known as elliptic 
 sin functions \cite{DO} (see also \cite{S} for various relevant identities). 
This provides another example of special
functions of a new kind related with quantum field theories.
Finally, in a recent paper \cite{W3} a
particular class of correlation functions in topological gauge field 
theories on non-compact
four-manifolds was proposed 
partially overlapping with the discussions in the second part of
this note.

{\em Acknowledgments}: The research was supported by  Grant
RFBR-09-01-93108-NCNIL-a. The research of AG was  also partly
supported by Science Foundation Ireland grant.

\section{Kurokawa multiple $L$-factors via topological field theory}

In \cite{GLO2} a representation of local Archimedean $L$-factors
as correlation functions  in equivariant
topological sigma models  on a disk
with target spaces $\IC^{\ell+1}$ was constructed. The local
Archimedean $L$-factors are basically given by products of
$\Gamma$-functions and \cite{GLO2} uses a realization of the
$\Gamma$-function as an inverse of a regularized infinite-dimensional determinant
obtained by taking an infinite-dimensional Gaussian integral. This
representation can be straightforwardly generalized
to higher dimensions. Multiple
$\Gamma$-functions introduced by Barnes \cite{Ba} allow 
a representation as inverse
regularized infinite-dimensional determinants. Below we recast this
representation into a framework of higher-dimensional topological
 equivariant field theories. Kurokawa proposed to use multiple $\Gamma$-functions
for construction of higher analogs of $L$-factors
and rise the question of their arithmetic interpretation \cite{Ku1},
\cite{Ku2} (see also \cite{Ma}). Thus the topological 
field theory representation of multiple $\Gamma$-functions provides a
topological  field theory representation of higher local $L$-factors
generalizing \cite{GLO2} to higher dimensions.   

We expect that the mirror dual representation of \cite{GLO3} 
has also a generalization to higher dimensions.  This provides an
instance of a higher dimensional analog of the 
 local Archimedean Langlands correspondence. In this Section we only briefly
touch this topic by calculating the multiple $\Gamma$-function via fixed
point localization of the corresponding topological field theory
integral (as it was demonstrated in \cite{GLO3} this directly leads to
mirror dual description for classical $\Gamma$-functions). 

\subsection{Multiple Gamma-functions}

Let us first recall the basic constructions of the hierarchy of
$\Gamma$-functions \cite{Ba}. The simplest $\Gamma$-function 
 (called elementary $\Gamma$-function in \cite{GL1}) is given by 
\be\label{G0}
\Gamma_0(s)=\frac{1}{s}.
\ee
The standard $\Gamma$-function can be considered as a next 
element of the hierarchy of $\Gamma$-functions. Actually it is more
natural to introduce  a slightly modified $\Gamma$-function
\be
\Gamma_{1}(s|\omega)=(2\pi)^{-\frac{1}{2}}\,\omega^{\frac{s}{\omega}-\frac{1}{2}}
\Gamma(s/\omega), \qquad \omega>0, 
\ee
satisfying  the functional equation
\be\label{mFF}
\frac{\Gamma_1(s|\omega)}{\Gamma_1(s+\omega|\omega)}=\Gamma_0(s). 
\ee
The modified $\Gamma$-function 
can be also expressed through a regularized infinite product 
\be
\Gamma_1(s|\omega)=\left[\prod_{n\in \IZ_{\geq 0}}
(s+n\omega)^{-1}\right]_{reg}:=\exp(\frac{\pr}{\pr \nu}\,
\zeta(s,\nu|\omega)|_{\nu=0},
\ee
where $\zeta(s,\nu|\omega)$ is defined as an analytic continuation of the
infinite sum 
\be
\zeta(s,\nu|\omega)=\sum_{n\in \IZ_{\geq}}\frac{1}
{(s+n\omega)^{\nu}}, \qquad {\rm Re}(\nu)>1. 
\ee
Another useful way to define (modified) $\Gamma$-function is  via the
Gauss integral representation of its logarithm 
\be\label{GaussR}
\ln
\Gamma_1(s|\omega)=-\gamma\left(\frac{s}{\omega}-\frac{1}{2}\right)+
\int_{\CC}\frac{dt}{2\pi\imath t} \frac{e^{-st}\ln(-t)}
{(1-e^{-\omega t})},
\ee
where  $\gamma$ is the Euler constant and 
 $\CC$ is the Hankel contour starting  at $+\infty$ enclosing $t=0$
 counterclockwise and returning to  $+\infty$.  

The elementary $\Gamma$-function \eqref{G0} can be obtained as a limit of the
modified $\Gamma$-function 
\be
\Gamma_0(s)=\lim_{\omega\to \infty}\,\left(\frac{\omega}{2\pi}\right)^{-\frac{1}{2}}\,
\,\Gamma_1(s|\omega), 
\ee
taking into account \eqref{mFF} and
$\Gamma_1(\omega|\omega)=(\omega/2\pi)^{1/2}$. 
Let $\underline{\omega}=(\omega_1,\ldots , \omega_r)$ be  an array of
positive real numbers. 
The multiple $\Gamma$-functions are defined  as  infinite products
\be\nonumber 
\Gamma_r(s|\underline{\omega})=\left[\prod_{\underline{n}\in \IZ^r_{\geq 0}}
(s+\<\underline{n},\underline{\omega}\>)^{-1}\right]_{reg}
=\frac{\pr}{\pr \nu}\,
\zeta_r(s,\nu|\underline{\omega})|_{\nu=0},\qquad
\underline{\omega}=(\omega_1,\ldots ,\omega_r),\quad \underline{n}=(n_1,\ldots
,n_r), 
\ee
regularized using analytic continuation of the 
$\zeta$-function \cite{Ba}
\be
\zeta_r(s,\nu|\underline{\omega})=\sum_{\underline{n}\in \IZ^{r}_{\geq}}\frac{1}
{(s+\<\underline{n},\underline{\omega}\>)^{\nu}}, \qquad {\rm Re}(\nu) >r.
\ee
Thus defined multiple $\Gamma$-functions satisfy the following defining
equations: 
 \be\label{FRbasic}
\frac{\Gamma_r(s|\underline{\omega})}{
\Gamma_r(s+\omega_i|\underline{\omega})}=
\Gamma_{r-1}(s|\underline{\omega}-\{\omega_i\}), \qquad i=1,\ldots, r,
\ee
and more generally 
\be\label{FR}
\prod_{\underline{\epsilon}} \Gamma_r(s+\<\underline{\omega},\underline{\epsilon}\>|
\underline{\omega})^{(-1)^{|\underline{\epsilon}|}}=s^{-1},
\ee
where $\underline{\epsilon}=(\epsilon_1,\ldots ,\epsilon_r)$, $\epsilon_i=0,1$
and $|\underline{\epsilon}|=\sum_i \epsilon_i$.

Multiple  $\Gamma$-functions allow integral representation
generalizing the classical Gauss integral representation 
\eqref{GaussR}
\be\label{GF}
\ln \Gamma_r(s|\underline{\omega})=
\frac{(-1)^r}{r!}\,\gamma\,B_{r,r}(s)+\,\,
\int_{\CC}\frac{dt}{2\pi\imath t} \frac{e^{-st}\ln(-t)}
{\prod_{a=1}^r(1-e^{-\omega_a t})},
\ee
where  $\CC$ is the Hankel contour. Here $\gamma$ is the Euler constant and
the  Bernoulli polynomial $B_{r,k}(s)$ are defined by the following
generating series:  
\be
\frac{t^re^{zt}}{\prod_{k=1}^{r}(e^{\omega_kt}-1)}=
\sum_{k=0}^{+\infty} B_{r,k}(z,\underline{\omega})\frac{t^k}{k!}. 
\ee
The following limiting relation between multiple
$\Gamma$-functions holds: 
\be\label{LIMIT}
\Gamma_{r-1}(s|\underline{\omega}-\{\omega_r\})=\lim_{\omega_r\to
  \infty}\,\frac{\Gamma_r(s|\underline{\omega})}
{\Gamma_r(\omega_r|\underline{\omega})}.
\ee
Let us note that hierarchy of multiple $\Gamma$-functions allows 
a natural $q$-deformation. The $q$-deformed multiple $\Gamma$-function 
is defined as an infinite product  
\be\label{qDEF}
\Gamma_r(t|\underline{q})=\prod_{\underline{n}\in \IZ^r_{\geq
    0}}\frac{1}{(1-tq_1^{n_1}\cdots q_r^{n_r})},
\ee
where $\underline{q}=(q_1,\ldots, q_r)$ and we imply $|q_j|<1$,
$j=1,\ldots ,r$. Under these conditions the product absolutely
converges and no regularization is needed. 
The $q$-deformed  multiple $\Gamma$-functions 
can be also represented as an infinite determinant
\be\label{qDEFd}
\Gamma_r(t|\underline{q})=\det_{\IC[z_1,\ldots, z_r]} 
\frac{1}{(1-tq_1^{d_1}\cdots q_r^{d_r})},
\ee
where the mutually commuting operators $d_j$  act as follows: 
\be
d_j\,\cdot z_1^{n_1}\cdots z_r^{n_r}= n_j\cdot z_1^{n_1}\cdots
z_r^{n_r}.
\ee 
One can  represent $q$-deformed multiple
$\Gamma$-functions as traces using the standard algebraic identities. 
Let $\CA_r$ be the space of polynomials 
of the variable $Z_{i,n}$, $i=1,\ldots ,r$, $n\in \IZ_\geq 0$.  
Then we have the following representation: 
\be
\Gamma_r(t|\underline{q})=\Tr_{\CA_r}\,t^Dq_1^{d_1}\cdots q_r^{d_r},
\ee
where the actions of the mutually commuting 
operators $D$ and $d_1,\ldots, d_r$ 
on $\CA_r$ are  induced by the following actions on the generators
of the polynomial ring:   
\be
d_j\,\cdot Z_{i,n}= \delta_{ij} n\, Z_{in}
,\qquad D\cdot Z_{jn}=Z_{jn}.
\ee
The multiple $q$-deformed 
$\Gamma$-functions satisfy $q$-analogs of the functional relations
\eqref{FR}
\be\label{qFR}
\prod_{\underline{\epsilon}} \Gamma_r(t \prod_{i=1}^rq_i^{\epsilon_i}|
\underline{q})^{(-1)^{|\underline{\epsilon}|}}=\frac{1}{1-t}. 
\ee
These  functional relations \eqref{qFR}  allow  a
simple interpretation in terms of coherent sheaves on
$\IC^r$. Consider for example the following exact sequence of 
$\IC[z]$-modules:
\be
0\longrightarrow z\IC[z]\longrightarrow \IC[z]\longrightarrow \IC \longrightarrow 0.
\ee
Using multiplicative  property of 
determinants with respect to exact sequences we have 
\be
\det_{\IC}(1-t)\,\det_{z\IC[z]}(1-tq^{d})=
\det_{\IC[z]}(1-tq^{d}),
\ee
where $d\cdot z^n=n\,z^n$. Taking into account the
definition \eqref{qDEF} of $q$-deformed $\Gamma$-functions we have 
the functional relation
\be
\frac{\Gamma_1(t|q)}{\Gamma_1(qt|q)}=\Gamma_0(t),
\ee
where 
\be
\Gamma_0(t)=\frac{1}{1-t}, \qquad \Gamma_1(t|q)=\frac{1}{\prod_{n=0}^{\infty}(1-tq^n)}.  
\ee
For $r=2$  consider  the Koszul resolution of the skyscraper sheaf
$\CO_{z=0}$ at $0\in \IC^2$
$$
0\longrightarrow \IC[z_1,z_2]\longrightarrow z_1\IC[z_1,z_2]\oplus z_2\IC[z_1,z_2]
\longrightarrow \IC[z_1,z_2] \longrightarrow \IC\longrightarrow 0.
$$
Let $d_1$, $d_2$ be commuting operators acting on polynomials 
via $d_i\cdot z_j^{n_j}=\delta_{ij}n_j\,z_j^{n_j}$. 
Taking determinant of the operator $(1-tq_1^{d_1}q_2^{d_2})$ acting on
various terms of the exact sequence we arrive at the following functional relation:
\be
\frac{\Gamma_2(tq_1q_2|q_1,q_2)\Gamma_2(t|q_1,q_2)}
{\Gamma_2(tq_1|q_1,q_2)\Gamma_2(tq_2|q_1,q_2)}=\frac{1}{1-t},
\ee
which is a particular instance  of \eqref{qFR}.  

\subsection{Topological field theory interpretation}

Now we  provide a simple interpretation of the
multiple $\Gamma$-functions (and thus Kurokawa generalized
local Archimedean $L$-factors)  as correlation functions 
in equivariant topological field theories thus generalizing   
the results of \cite{GLO2} to higher dimensions. We start with a
representation of multiple $\Gamma$-functions as properly regularized
symplectic volumes of infinite-dimensional spaces. Consider 
the  space $\CM(D^r,\IC^{\ell+1})$
of holomorphic maps of the polydisk  $D_r=\{(z_1,\ldots
,z_r)\in \IC^r|\,\,|z_a|\leq 1\}$  into the  vector space  $\IC^{\ell+1}$. There are
natural actions of $U_{\ell+1}$ and  real $r$-dimensional torus  
$T_r=S^1\times \cdots \times S^1$ on $\CM(D^r,\IC^{\ell+1})$. 
The action of $U_{\ell+1}$ is induced from the standard action on $\IC^{\ell+1}$
and $T_r$ acts by rotation of  $D_r$
\be
z_a\rightarrow e^{\imath \alpha_a}z_a,\qquad a=1,\ldots ,r.
\ee
The space of maps has natural symplectic structure
\be\label{sympform}
\Omega=\frac{\imath}{2(2\pi)^r}\,
\int_{|z_a|=1}\prod_{a=1}^{r}d\sigma_a\,\,\,\sum_{j=1}^{\ell+1}\delta
\varphi^j(z)\wedge \delta \bar{\varphi}^j(z),\qquad 
z_a=r_ae^{\imath \sigma_a}, 
\ee
induced by the standard symplectic structure on $\IC^{\ell+1}$
\be\label{stss}
\omega=\frac{\imath}{2}\sum_{j=1}^r dz_j\wedge d\zb_j.
\ee
The action of $U_{\ell+1}\times T_r$ is Hamiltonian with respect to 
\eqref{sympform} and the
corresponding  momenta for the  diagonal subgroup
$U_1^{\ell+1}\subset U_{\ell+1}$ and $T_r$ are given by
\be
H_b=\frac{\imath}{2(2\pi)^r}\,\,\,
\int_{|z_a|=1}\prod_{a=1}^r d\sigma_a\,\,\,\sum_{j=1}^{\ell+1}\bar{\varphi}^j
\pr_{\sigma_b}\varphi^j,\qquad b=1,\ldots ,r,
\ee
\be
H_j=-\frac{1}{2(2\pi)^r}\,
\int_{|z_a|=1}\prod_{a=1}^r d\sigma_a\,\,\,|\varphi^j|^2,\qquad j=1,\ldots ,(\ell+1).
\ee
We define $T_r\times U_{\ell+1}$-equivariant volume of 
$\CM(D_r,\IC^{\ell+1})$ as the following 
infinite-dimensional integral:
\be\label{eqvol}
Z(\underline{\lambda},\underline{\omega})
=\int_{\CM(D_r,\IC^{\ell+1})}\,\,e^{\Omega-\sum_{j=1}^{\ell+1}
 \lambda_jH_j -\sum_{a=1}^r\omega_aH_a},\qquad \omega_a>0
\ee
where $\underline{\lambda}=(\lambda_1,\ldots,\lambda_{\ell+1})$, 
$\underline{\omega}=(\omega_1,\ldots,\omega_{\ell+1})$ and  
the infinite-dimensional 
Gaussian integrals  are understood via $\zeta$-function regularization.
Straightforward calculations similar to the one in \cite{GLO2} give
\be\label{mG}
Z(\underline{\lambda},\underline{\omega})=\prod_{j=1}^{\ell+1}
\Gamma_r(\lambda_j|\underline{\omega}).
\ee
It is easy to write down representation for the
equivariant symplectic volume \eqref{eqvol} using formalism of topological field
theories. Consider a quantum field theory on $D_r$
 with the following set of fields:
\be\label{FC}
(\varphi^j,\bar{\varphi}^j,\chi^j,\bar{\chi}^j,
\psi^{ja},\bar{\psi}^{ja},F^{ja},\bar{F}^{ja}),
\qquad j=1,\ldots (\ell+1),\quad a=1,\ldots r,
\ee
where $\varphi$ and $F$ are even and $\psi$ and $\chi$ are
odd fields. Note that the complex functions  
$(\varphi,F,\psi,\chi)$ can be written in terms of real
functions 
\be
(\varphi^A,\chi^A,\psi^{A\mu},F^{A\mu}),
\qquad A=1,\ldots 2(\ell+1),\quad \mu=1,\ldots 2r, 
\ee
with additional constraints
\be\label{constrour}
\psi_{\mu}^A+(J_2)_{\mu}^{\nu}\psi_{\nu}^B(J_1)_B^A=0,\qquad
F_{\mu}^A+(J_2)_{\mu}^{\nu}F_{\nu}^B(J_1)_B^A=0.
\ee
Here  $J_1\in {\rm End}(\IR^{2r})$ and $J_2\in {\rm
  End}(\IR^{2\ell+2})$ are standard complex structures on $\IC^r$ and
$\IC^{\ell+1}$ correspondingly.

Let us pick a $T_r$-invariant metric $h$ on $D_r$. 
Consider a topological theory with 
 the following action functional:  
\be\label{AF}
S_0=\delta_0 \left(\int _{D^r}d^{2r}z\,\sqrt{h}\CV\right)=
\int _{D^r}d^{2r}z\,\sqrt{h}h^{\mu\nu}(F_{\mu}^A\pr_{\nu}\phi^A+
\psi_{\mu}^A\pr_{\nu}\chi^A),
\ee
where
\be
\CV=h^{\mu\nu}\psi_{\mu}^A\pr_{\nu}\phi^A,
\ee
and $\delta_0$ is the  BRST operator defined by the relations 
\be\label{BRST}
\delta_0 \phi^A=\chi^A,\qquad \delta_0 \chi^A=0,\qquad
\delta_0 \psi_{\mu}^A=F_{\mu}^A,\qquad \delta_0 F_{\mu}^A=0.
\ee
The operator $\delta_0$ is obviously nilpotent $\delta_0^2=0$.  

We are interested in construction of $U_{\ell+1}\times
T_r$-equivariant  generalization of the theory with the action
\eqref{AF}. Let us start with $T_r$-equivariance.  
The $T_r$-equivariant differential is given by standard equivariant
extension of the de Rham differential   
\be
\delta_{T_r}=\delta_0+\sum_{a=1}^{r}\omega_a\iota_{v_a},\qquad 
v_a=\imath \left(z_a\frac{\pr}{\pr z_a}-\zb_a\frac{\pr}{\pr
    \zb_a}\right), 
\ee
and $(\omega_1,\ldots,\omega_r)$ is an element of the Lie algebra
${\rm Lie}(T_r)$. Consider the action of the diagonal subgroup 
$U(1)^{\ell+1}$ on $\IC^{\ell+1}$ given by
\be
e^{\imath \alpha_j}:\,\,\,\varphi^k\longrightarrow e^{\imath
  \alpha_j\delta_{j,k}}
\varphi^k,\qquad j,k=1,\ldots, (\ell+1).
\ee
Now $U_1^{\ell+1}\times T_r$-equivariant generalization of the BRST
operator \eqref{BRST} is given by
\be\label{eBRST}
\delta\varphi^j=\chi^j,\qquad \delta
\chi^j=-\left(\sum_{a=1}^r\omega_ai_{v_a} d\varphi^j
+\imath \lambda_j\varphi^j\right),
\ee
\be\nonumber
\delta \psi^j=F^j,\qquad \delta
F^j=-\left(\sum_{a=1}^r\omega_a\CL_{v_a}\psi^j+\imath \lambda_j\psi^j\right).
\ee
where $(\lambda_1,\ldots, \lambda_{\ell+1})$ is an element of Lie
algebra ${\rm   Lie}(U_1^{\ell+1})$. 
Consider the following $U_1^{\ell+1}\times
T_r$-invariant $\delta$-closed form on the space of maps
$D_r\to \IC^{\ell+1}$: 
\be\label{eqO}
\CO=\frac{\imath}{2(2\pi)^r}\int_{|z_a|=1}\,\prod_{a=1}^rd\sigma_a\,\,\,
(\sum_{j=1}^{\ell+1}\chi^j\bar{\chi}^j+\imath \lambda_j|\varphi^j|^2
+\sum_{a=1}^r\omega_a\bar{\varphi}^j\pr_{\sigma_a}\varphi^j).
\ee
Now we would like to calculate the functional integral with the
following modified action
\be\label{actful}
S=\int_{D^r}d^{2r}z\,\sqrt{h}h^{\mu\nu}(F_{\mu}^A\pr_{\nu}\phi^A+
\psi_{\mu}^A\pr_{\nu}\chi^A)+\CO.
\ee
Integrating over $F$ we restrict the fields $\phi$ to the
subspace
\be
\apr \phi^i=0,\qquad \pr \bar{\phi}^i=0, 
\ee
and similarly for integration over $\psi$.  
Further integral over holomorphic fields $\phi$ and $\chi$ 
reduces to the equivariant volume integral \eqref{eqvol}.  
Thus the functional  integral with the action \eqref{actful}
gives the integral representation for multiple $\Gamma$-function 
\be\label{Th}
Z(\underline{\lambda}|\underline{\omega})
=\prod_{j=1}^{\ell+1}\Gamma_r(\lambda_j|\underline{\omega}).
\ee
Let us also  note that the topological field theory
with the action \eqref{actful} for $r=2$ and $\ell+1=2$
can be interpreted as an $U_2\times T_2$-equivariant 
extension  of the theory of $\CN=2$ SUSY
hypermultiplet. 

Recall that classical $\Gamma$-function provides a basic building block 
of the Mellin-Barnes integral
representation  of various special functions \cite{KL}, \cite{GKL}. 
Given a quantum field theory
representation of multiple $\Gamma$-functions, 
one might  expect that the special functions allowing
the Mellin-Barnes integral representations  have
natural  higher-dimensional generalizations  expressed in terms of multiple
$\Gamma$-functions and related with 
correlation functions in higher-dimensional equivariant 
topological field theories.

\subsection{Multiple Euler integrals via  fixed point calculation}

In the case of two-dimensional disk (i.e. for $r=1$) the functional integral in
the  left hand side of \eqref{Th} was interpreted as a correlation
function in type $A$ 
topological sigma model i.e. as a topological  sigma model obtained
by type $A$ twisting from a $\CN=2$ SUSY sigma model \cite{GLO1}. One
can consider a mirror dual type $B$ topological sigma model  of the Landau-Ginzburg
type. This dual description naturally leads to 
the Euler integral representation of the 
 classical $\Gamma$-function. In \cite{GLO3} it was demonstrated that the type $B$
 dual description can be obtained directly from the type $A$ functional integral 
using fixed point localization technique. Below we provide a heuristic 
derivation of the multiple analog of the Euler integral representation 
of multiple $\Gamma$-functions applying fixed point localization to the
infinite-dimensional integral \eqref{eqvol}. Thus obtained integral
representations of multiple $\Gamma$-functions shall naturally arise
in mirror dual description of the topological field  theories with the
actions \eqref{actful}. 

Let us first  briefly recall the derivation of the Euler integral 
representation of the classical $\Gamma$-function via fixed point
localization \cite{GLO3}. Precisely we are going to derive an expression for
 the Fourier transform of  $\Gamma$-function (the Euler integral
representation  is an inversion of this representation)
\be\label{mEULER}
E_1(\tau|\omega):=\frac{1}{2\pi \imath}
\int_{-\imath \infty +\epsilon}^{+\imath \infty +\epsilon}
\,ds\,e^{-s \tau}\,\Gamma_1(s|\omega)=\left(\frac{\omega}{2\pi}\right)^{\frac{1}{2}}
\,e^{-\frac{1}{\omega}e^{\omega\tau}}. 
\ee
The standard functional equation on $\Gamma$-function can be written
as the following  equation on its Fourier transform
\be\label{Eeq}
(\pr_{\tau}+e^{\omega\tau})E_1(\tau|\omega)=0.
\ee
Recall that $\Gamma$-function allows a representation as
the  infinite-dimensional integral \eqref{eqvol}, \eqref{mG}. 
The Fourier transform of the integral \eqref{eqvol} gives a
representation of the $U_1\times S^1$- equivariant  volume of
the projectivization $\IP(\CM(D_1,\IC))$ of the space of
holomorphic maps $\CM(D_1,\IC)$ of the disk $D_1$ into the complex plane. 
This symplectic space allows Hamiltonian action of $S^1$  by
rotations of $D_1$. The $S^1$-fixed points in $\IP(\CM(D_1,\IC))$
 in homogeneous coordinates (given by coefficients of
series expansions $\varphi(z)=\sum_{m=0}^{\infty}\varphi_mz^m$ of the
holomorphic maps $\varphi:\,D_1\to \IC$) are enumerated by $n\in
\IZ_{\geq 0}$ and given by 
\be
\varphi_m=0, \qquad m\neq n. 
\ee 
Local coordinates near $n$-th isolated fixed point are then can be
chosen as $\varphi_m/\varphi_n$, $m\neq n$. Formal application of the 
Duistermaat-Heckman formula \cite{DH} gives the following heuristic expression
for the Fourier transform 
\be
\frac{1}{2\pi \imath}\int_{-\imath \infty +\epsilon}^{+\imath \infty +\epsilon}
\,d\lambda\,e^{-\lambda \tau}\,Z(\lambda,\omega)=
\sum_{n=0}^{\infty}\frac{\prod_{m=1}^{\infty}
(m\omega)}{\prod_{m=0,\,m\neq n}^{\infty}
((m-n)\omega)}\,\,e^{S_n},\qquad S_n=n\omega \tau,
\ee  
where we normalize the right hand side by multiplying on  
$\tau$- and $n$-independent function $\prod_{m=1}^{\infty}
(m\omega)$. The infinite product above is  understood via
$\zeta$-function regularization.  Now formal manipulations give  
\be
\frac{1}{2\pi \imath}\int_{-\imath \infty +\epsilon}^{+\imath \infty +\epsilon}
\,d\lambda\,e^{-\lambda \tau}\,Z(\lambda,\omega)=
\left(\frac{\omega}{2\pi}\right)^{\frac{1}{2}}\,
\sum_{n=0}^{\infty}\frac{(-1)^n}{n!\omega^n}\,e^{n\omega
  \tau}=\left(\frac{\omega}{2\pi}\right)^{\frac{1}{2}}
e^{-\frac{1}{\omega}e^{\omega \tau}}, 
\ee
Thus taking into account the relation $Z(\lambda,\omega)=\Gamma_1(\lambda|\omega)$
we arrive at \eqref{mEULER}. It is useful to rewrite  the series
expansion for $E_1(\tau|\omega)=(\omega/2\pi)^{1/2}\,
\exp(-\frac{1}{\omega}e^{\omega \tau})$
in the following form 
\be\label{serE}
E_1(\tau|\omega)=\left(\frac{\omega}{2\pi}\right)^{\frac{1}{2}}\,
\sum_{n=0}^{\infty}\,\,\left(\frac{\Gamma_1(\epsilon-n\omega|\omega)}
{\Gamma_1(\epsilon|\omega)}e^{(n\omega-\epsilon)\tau}\right)_{\epsilon\to+ 0}. 
\ee
It is easy to verify that the equation \eqref{Eeq} reduces to  the
basic functional relation \eqref{mFF} on coefficients of the series \eqref{serE}.

Now we apply similar approach to  higher-dimensional integrals
\eqref{eqvol} to obtain multiple analogs of \eqref{mEULER} and
\eqref{serE}.  Consider Fourier transformed  infinite-dimensional
integral \eqref{eqvol} for  $\ell=0$ and arbitrary $r$
\be\label{mFT}
E_r(\tau|\underline{\omega})=\frac{1}{2\pi \imath}
\int_{-\imath \infty +\epsilon}^{+\imath \infty +\epsilon}
\,d\lambda\,e^{-\lambda \tau}\,Z(\lambda,\underline{\omega}),
\ee
where $\underline{\omega}=(\omega_1,\ldots, \omega_r)$. 
We would like to calculate \eqref{mFT}  explicitly using fixed point
localization with respect to the action of the torus $T_r$. 
By \eqref{mG} this gives an integral transform of multiple
$\Gamma$-function which can be inverted to
obtain multiple analog of the Euler integral representation    
\be
\Gamma_1(s|\omega)=\left(\frac{\omega}{2\pi}\right)^{\frac{1}{2}}\,
\int_{-\infty}^{+\infty}\,d\tau\,e^{s\tau}\,e^{-\frac{1}{\omega}
e^{\omega \tau}}. 
\ee
Analogously to the case of $r=1$ (see \cite{GLO3}) the set of fixed
points of $T_r=S^1\times\cdots \times S^1$ on $\IP(\CM(D_r,\IC))$ is 
enumerated by $r$-tuples $\underline{n}=(n_1,\ldots ,n_r)$
of non-negative integers. The local
coordinates near isolated fixed points are given by 
$\varphi_{\underline{m}}/\varphi_{\underline{n}}$,
$\underline{m}=(m_1,\ldots ,m_r)\neq \underline{n}=(n_1,\ldots ,n_r)$. 
The sum over fixed points boils down to the following: 
\be
E_r(\tau|\underline{\omega})=
\sum_{\underline{n}\in \IZ_{\geq 0}^r}\frac{\prod_{\underline{m}\in
    (\IZ^r_{\geq 0}-\{0\})}
\<\underline{m},\underline{\omega}\>}{
\prod_{\underline{m}\in  (\IZ^r_{\geq 0}-\{\underline{n}\})}
\<\underline{m}-\underline{n},\underline{\omega}\>}
\,\,e^{S_{\underline{n}}},\qquad S_{\underline{n}}=\<\underline{n},
\underline{\omega}\>\,\tau. 
\ee  
This formal expression can be rewritten in terms
higher $\Gamma$-functions  
\be\label{mserE}
E_r(\tau|\underline{\omega})=
E_r(0|\underline{\omega})\,
\sum_{\underline{n}\in \IZ_{\geq 0}^r}
\left(\frac{\Gamma_r(\epsilon-\<\underline{n},\underline{\omega}\>|\underline{\omega})}
{\Gamma_r(\epsilon|\underline{\omega})}\,\,
e^{(\<\underline{n},\underline{\omega}\>-\epsilon)\,\tau}\right)_{\epsilon\to +
0}. 
\ee
On the other hand by taking into account \eqref{mG} and \eqref{mFT} we obtain 
the (Fourier transformed) multiple analog of the Euler integral
representation  for multiple $\Gamma$-functions.  Note that it is easy to check
that the series \eqref{mserE} satisfies the Fourier transformed analog
of the basic functional relation \eqref{FRbasic}
\be
\left(\frac{1}{\Gamma_{r-1}(\pr_\tau|\underline{\omega}-\{\omega_a\})}+e^{\omega_a\tau}
\right)\,E_r(\tau|\underline{\omega})=0,\qquad a=1,\ldots ,r.
\ee
Our considerations in this Subsection were mostly heuristic and
 additional work is obviously needed to make these reasonings  precise.  

\section{Topological gauge field theories in $d=2$ and $d=4$}

In \cite{GLO4}  the results of \cite{GLO2}, \cite{GLO3}
were generalized to the case of  two-dimensional type $A$ topological sigma 
models with the compact target spaces $\IP^{\ell}$. We identify 
a particular correlation function in type $A$ topological
sigma model on a disk $D$ with 
 the  Whittaker function associated  with the  maximal parabolic
subgroup of $GL_{\ell+1}$. This provides a 
 representation of the Whittaker functions as an  
infinite-dimensional integral the over space of holomorphic maps of $D$ in
 $\IP^{\ell}$. In explicit calculation of the corresponding correlation function 
the  representation of the topological sigma model with the target space
$\IP^{\ell}$ via a linear $U(1)$-gauged sigma model with the target
space $\IC^{\ell+1}$ was used.
In \cite{GLO4} we also give a mirror dual description
of the  type $A$ twisted  topological $\IP^{\ell}$-sigma model 
in terms of a type $B$ twisted Landau-Ginzburg model  reproducing
the  finite-dimensional integral representation of the Whittaker
function \cite{GKLO}. One shall stress that the correlation functions
in type $A$ topological $\IP^{\ell}$-sigma model on a disk are closely connected
 with the counting two-dimensional instantons i.e. 
 holomorphic maps of $\IP^1$ in  $\IP^{\ell}$ \cite{Gi1}, \cite{Gi2},
 \cite{GLO1} (the Whittaker functions   appear already  in 
instanton counting \cite{Gi3}). However the  most direct
connection between the Whittaker functions  and  correlation functions
in  two-dimensional topological quantum field theories arises when the
latter are considered on two-dimensional disks.  

To discuss further generalizations of the results of  \cite{GLO4} 
let us note that the calculation of a particular correlation function in topological
$\IP^{\ell}$-sigma model \cite{GLO4} 
can be reduced to a calculation
of $S^1\times U_{\ell+1}$-equivariant symplectic volume of the space
of holomorphic maps of the  disk $D$ into $\IP^{\ell}$. This
formulation has obvious relation with the basic setup of the
Floer cohomology groups of Lagrangian submanifolds calculated via counting holomorophic
disks. There is a well-known four-dimensional  analog \cite{F} (see
also e.g. \cite{CJS}, \cite{AB} and \cite{DK} for general facts on
instanton moduli spaces) of the theory where the role of the space of
holomorphic maps of two-dimensional disks into symplectic manifolds
is played by the moduli spaces of instantons i.e. 
the spaces of gauge equivalence classes of
 anti self-dual connections on principle
$G$-bundles over  four-dimensional manifolds with nontrivial
boundaries. The cohomology invariants of the instanton moduli spaces 
(such as e.g. equivariant symplectic volumes) can be conveniently described in terms of
topological Yang-Mills theories  \cite{W1}. Thus one would expect that
a generalization of the results of \cite{GLO4} to  the case of
topological four-dimensional Yang-Mills gauge theories
would provide an interesting example of higher analog of
special function.  Let us remark  that in four dimensions there is an
analog of two-dimensional mirror-symmetry known as $S$-duality. In the case of
asymptotically free $\CN=2$ Yang-Mills theories with a gauge group
$G$  the $S$-dual theory is a
$\CN=2$ Yang-Mills  theory with an abelian gauge group dual to abelian subgroup
of the original gauge group $G$  interacting with monopole
hypermultiplets. There exists an explicit 
construction of the mirror duals to 
two-dimensional  gauged linear sigma-models \cite{AV}. 
Thus one might  hope  that there is a similar  effective dual
description of the correlation functions 
in the topological non-abelian Yang-Mills theories via 
$S$-dual topological theories. 
Let us stress that the explicit
 calculations of a particular instanton counting functions in $\CN=2$ SUSY Yang-Mills
theories  were initiated  in \cite{LNS} and
its relation with Seiberg-Witten geometry was demonstrated in
\cite{N}, \cite{NO}, \cite{NY} by explicit calculations
(see also \cite{BE} where the affine Whittaker
functions were related with instanton counting in topological
Yang-Mills theories). 
Thus pursuing the analogy with \cite{GLO4} in the four-dimensional
case one may hope to find a more direct and conceptional explanation of the
relation between instanton counting and Seiberg-Witten solution of
$\CN=2$ SUSY Yang-Mills theory.
In this part of the note we briefly describe  basic constructions
in  topological field  theories on two- and four-dimensional
non-compact  manifolds relevant  to the program of deriving dual  pairs of
integral representations of new special functions associated with
topological gauge field theories in four dimensions. We postpone  detailed 
considerations including  more general examples for the future publication \cite{GL2}.

\subsection{Equivariant symplectic volumes of
  instanton moduli spaces}

Let us first recall basic constructions used in \cite{GLO4} with the
emphasis on a relation with the Floer cohomology theory
(see e.g. \cite{CJS}, \cite{Gi1}).

Let $X$ be a K\"{a}hler manifold with a K\"{a}hler form
$\omega$. Let  $\widetilde{LX}$ be a universal cover of the loop
space $LX$ of $X$. Consider a submanifold $\widetilde{LX}_+\subset \widetilde{LX}$
of the loops $\Phi:\,S^1\to X$ 
 allowing an extension to  holomorphic maps of the disk
$D$, $\pr D=S^1$ into $X$.  This is a sympelctic manifold with the sympelctic structure
\be\label{sympstr2}
\Omega_2=\int_{S^1}\,d\sigma\,\,
\omega_{i\jb}(\varphi,\bar{\varphi})\,
\delta \varphi^i\,\wedge \delta \bar{\varphi}^{\jb},
\ee
where $\Phi$ is locally described by a 
set of functions $(\varphi^i(\sigma),\bar{\varphi}^i(\sigma))$. 
The group $S^1$ of loop rotations acts on
$(\widetilde{LX}_+,\Omega_2)$ in a Hamiltonian way and we denote 
the corresponding momentum by $H_{S^1}$. 
Suppose that $X$ is also  supplied with the  Hamiltonian action of a compact Lie group
$G$ and $\mu(\varphi,\bar{\varphi})$ be  the corresponding momentum map $\mu: X\to
\mathfrak{g}^*$, $\Fg={\rm Lie}(G)$. The momenta for induced action of $G$ on
$\widetilde{LX}_+$  are then given by
\be
H_a=\int_{S^1}\,d\sigma\,\,
\mu_a(\varphi(\sigma),\bar{\varphi}(\sigma)), \qquad a=1,\ldots, {\rm dim}(G).
\ee
In \cite{GLO2}, \cite{GLO4}  the following $S^1\times G$-equivariant symplectic
volume integrals were considered
\be\label{eqvol2}
Z=\int_{\widetilde{LX}_+}\,e^{\Omega_2-\sum_{a}\lambda_a\,H_a-\hbar H_{S^1}},
\ee
where $\Omega_2-\sum_{a}\lambda_a\,H_a-\hbar
H_{S^1}$ shall be understood as  $S^1\times G$-equivariant extension
of the symplectic form \eqref{sympstr2}.
It was demonstrated in \cite{GLO2}, \cite{GLO4}  that such integrals 
for particular $(X,\omega,G)$ provide 
infinite-dimensional integral representations of special functions
such as $\Gamma$-function and various  Whittaker functions.

The construction described above allows  the following four-dimensional
generalization. Let $M$ be a four-dimensional manifold with a boundary
$N=\pr M$. One considers  the universal cover
$\widetilde{\CA_N/\CG}$ of the space $\CA_N/\CG$
of gauge equivalence classes of connections
on a principle $G$-bundle over a three-dimensional manifold $N$
(it might be more natural to consider the factorization over subgroup
of basic gauge transformations $\CG_0\subset\CG$ i.e. respecting a
trivialization at a point in $N$). 
Let $N$ allow an  action of an isometry group $G_N$ and this action can
be extended to $M$. The space $\widetilde{\CA_N/\CG}$ 
shall be considered as an analog of the space
$\widetilde{LX}$ in two-dimensional case and  
an appropriate  subgroup of $G_N$ will play the role 
of the group $S^1$ of disk rotations.  The analog of the space of
holomorphic maps $D\to X$ is then the space $\CM(M,N)$ of 
gauge equivalence classes of anti self-dual $G$-connections on $M$ 
understood as a subspace in $\widetilde{\CA_N/\CG}$ via restriction to
the boundary $N$. 
There is also an action of the group $G$ of global gauge transformations
on the moduli space $\CM(M,N)$ of anti-self-dual $G$-connections on
$M$. In the calculation  of the correlation function 
in $\IP^{\ell}$ topological sigma model \cite{GLO4}  
 the basic observable in the corresponding linear $U(1)$ gauged 
 sigma model on the disk was the integral $\CO=\int_D F(A)$
representing  the pull back of the standard 
symplectic form on the target space $\IP^{\ell}$. In
four-dimensional case the formal analog is given by the integral
\be
\CO=\int_M \Tr F(A)\wedge F(A),
\ee
where $F(A)$ is a curvature of the  connection $A$ on a principle
$G$-bundle on $M$. 

Let us consider a simple case of $N=S^1\times \Sigma$, 
$\Sigma$ being  a compact two-dimensional surface 
and $G_N=S^1$ acts on the first factor by rotations. 
The symplectic structure on $\CM(N,M)$ is given by a restriction of
the following two-form
\be\label{4SF}
\Omega_4=\frac{1}{2}\int_{N=S^1\times \Sigma}\,e_{S^1}\wedge \Tr\,\,
\delta A \wedge \delta A,
\ee
 where $e_{S^1}$ is a lift of constant one-form $d\theta$ on the first
 factor $S^1$ of $N=S^1\times \Sigma$. 

In analogy with \eqref{eqvol2}
we would like to calculate  $S^1\times G$-equivariant symplectic volume of
the space $\CM(M,N)$
\be\label{eqvol4}
Z(\phi_0,\hbar)=\int_{\CM(M,N)}\,\,\,e^{\Omega^{equiv}_4}, \qquad
\phi_0\in {\rm Lie}(G),
\ee
where $\Omega_4^{equiv}$ is following 
$S^1\times G$-equivariant extension of the symplectic form \eqref{4SF}:
\be
\Omega_4^{equiv}=\int_{N=S^1\times \Sigma}\,e_{S^1}\wedge \Tr
(\frac{1}{2}\delta A \wedge \delta A+\phi_0F(A))+\hbar S_{CS}(A),
\ee
and $S_{CS}(A)$ is the Chern-Simons functional
\be
S_{CS}(A)=\int_N\,\Tr(A\,d\,A+\frac{2}{3}A^3). 
\ee
This integral  basically reduces to the functional integral
in the  Chern-Simons theory on the boundary $N$ restricted to the
connections such  that the connections on $N$ 
can be extended to anti  self-dual connections  on  $M$.
Note that the integrand in \eqref{eqvol4} 
is not invariant with respect to
large gauge transformations (related with non-trivial instantons in
the bulk). This is consistent with the fact that we consider universal
cover $\widetilde{\CA_N/\CG_0}$ instead of $\CA_N/\CG_0$.
In the following Subsection we rewrite this  integral
using the standard formalizm of topological  gauge field theories \cite{W1}.

\subsection{Topological field theory representation of symplectic volumes}

Equivariant symplectic  volume \eqref{eqvol4} of the moduli spaces of
instantons $\CM(M,N)$ 
can be represented as the  functional integral in a topological gauge
field theories \cite{W1}. Let us recall basic construction of the
topological gauge theory associated with a principle $G$-bundle
$\CP_G$ over a four-dimensional manifold $M$. Let $\mathfrak{g}={\rm
  Lie}(G)$ be a Lie algebra of $G$. Topological gauge theory field multiplet
$(A,\psi,\phi)$ consists of a connection $A$,   
$\Fg$-valued odd one form $\psi$ and a $\Fg$-valued complex even scalar field
$\varphi$. More precisely the last two fields take values in
$ad_{\Fg}$-bundles associated with $\CP_G$. 
 The  BRST transformations are defined as follows:
\be
\delta A=\psi,\qquad \delta \psi=-D\phi,\qquad \delta \phi=0,
\ee
where $D\phi$ stands for  covariant derivative 
$D_{\mu}\phi=\pr_{\mu}\phi+[A_{\mu},\phi]$.
 Consider additional field multiplet consisting  of an anti self-dual two form
$\chi$
\be
\chi_{\mu\nu}^a=-\chi_{\nu\mu}^a=\frac{1}{2}\epsilon_{\mu\nu\rho\tau}\chi^{a\rho\tau},
\ee
its BRST partner $H$
\be
\delta \chi=H,\qquad \delta H=[\phi,H],
\ee
and a pair of even and odd  zero forms $\lambda$, $\eta$ 
\be
\delta \lambda=\eta,\qquad \delta \eta=[\phi,\lambda].
\ee
The action is given by a $\delta$-variation
\be\label{YM}
S=\delta\left(\int_M d^4x\,\, \CV\right)=
\int_M\,d^4x \Tr (-2H^{\mu \nu}F^+_{\mu\nu}+\frac{1}{2}\phi
  D_{\mu}D^{\mu}\lambda-\eta  D_{\mu}\psi^{\mu}
\ee
$$
 -\lambda [\psi_{\mu},\psi^{\mu}]-
\chi^{\mu\nu}(D_\mu\psi_\nu-D_{\nu}\psi_{\mu}-
\epsilon_{\mu\nu\rho\tau}D^{\rho}\psi^{\tau})),
$$
where $\CV=\Tr\left(-D_{\mu}\lambda
  \psi^{\mu}+2\chi^{\mu\nu}F^+_{\mu\nu})\right)$ and
$F^+_{\mu\nu}=\frac{1}{2}(F_{\mu\nu}+*F_{\mu\nu})$.
After integration over $H$  in the functional integral with the action
\eqref{YM} the integral over gauge connections is localized  on a
subset of anti self-dual connections $F^+=0$. Further integration over
$\chi$ and $\eta$ leads to the constraints
\be
D_\mu\psi_\nu-D_{\nu}\psi_{\mu}-\epsilon_{\mu\nu\rho\tau}D^{\rho}\psi^{\tau}=0,\qquad
D_{\mu}\psi^{\mu}=0.
\ee 
These constraints effectively restrict  $\psi$ to be a section of 
the tangent bundle to the moduli space of anit self-dual connections.
Thus the functional integral (modulo some subtleties with zero modes
\cite{W1}) reduces to the integral over odd tangent bundle to the
moduli space of instantons.  

Now consider an $S^1\times G$-equivariant extension of the  
topological Yang-Mills theory TYM theory on $D\times \Sigma$
 where $S^1$ acts by rotations of the disk $D$. 
The $S^1\times G$-equivariant BRST operator acts on the
topological  gauge multiplet as follows (compare with \cite{GLO4})
\be\label{gtr}
\delta_{eq}\,A= \psi,\qquad
\delta_{eq}\,\psi=-D\phi+\hbar d(\iota_{v} A)+\hbar \iota_{v}F(A),
\qquad \delta_{eq} \phi=0,
\ee
$$
\delta_{eq} \chi=H,\qquad \delta_{eq} H=\hbar\CL_{v}\chi,\qquad
\delta_{eq} \lambda=\eta,\qquad \delta_{eq} \eta=\hbar\CL_{v}\lambda,
$$
where $v$ is a generator of $S^1$.  
The four-manifold $M=D\times \Sigma$ has the boundary  $N=S^1\times
\Sigma$ and the following  observable is $\delta_{eq}$-closed
\be
\CO=\int_{S^1\times \Sigma}\,e_{S^1}\wedge \,\Tr (\frac{1}{2}\psi\wedge \psi+\phi
F(A))+\hbar S_{CS}(A),
\ee
where
\be
S_{CS}(A)=\int_{S^1\times
\Sigma}\,\Tr\,\left( AdA+\frac{2}{3}A^3\right),
\ee
is the  Chern-Simons functional. This is precisely the observable 
we use in \eqref{eqvol4} to define the  equivaraint symplectic volume 
of the moduli space of instantons on non-compact four-manifolds.
To calculate the integral \eqref{eqvol4} one can use 
equivariant localization or/and  explicit parametrization
of anti self-dual fields via twistor formalizm. Note that one can 
consider another interesting examples of four-manifolds with
non-trivial isometries such as $D\times D$ with a natural action of
$S^1\times S^1$ rotating two disks independently. This  leads to
a consideration of the Chern-Simons theory on $S^1\times D$ i.e. to
a potential connection with conformal field theories (compare with
\cite{AGT}).

\subsection{On a dual description of equivariant symplectic
  volumes}

General approach to study  topology of
instanton moduli spaces via  $S$-dual  quantum gauge field theories was proposed
in \cite{W2} and successfully applied to calculations of the Donaldson
invariants of compact four-dimensional manifolds. One would expect
that the same approach should work for the calculation of equivariant
volumes \eqref{eqvol4} on non-compact four-manifolds (see \cite{GW}
for related considerations). Let us stress that similar 
 approach based on mirror symmetry successfully works for
non-compact two-dimensional surfaces \cite{GLO4}. A mirror 
 dual description of two-dimensional sigma models with the target
 spaces being compact K\"{a}hler manifolds
 with positive first Chern class  
in terms of  Landau-Ginzburg theories   
 leads to explicit finite-dimensional
integral representations of the corresponding equivaraint symplectic volume integrals
 \eqref{eqvol2}.  

Let us provide some general remarks on analogy between four-dimensional $S$-duality
versus two-dimensional mirror symmetry relevant to calculations of
equivariant symplectic volumes.  The approach of \cite{W2} can be considered as an
application of $S$-duality in the
following sense.  Recall that $S$-duality
 transformation of abelain gauge fields  relates  
on-shell gauge field $A$ and its dual $A^{\vee}$ via the
constraint $F(A^{\vee})=*F(A)$. This relation can be non-trivially generalized
to finite non-abelian theories such as  $\CN=4$ SUSY Yang-Mills
theories or finite $\CN=2$ SUSY $SU(N)$ Yang-Mills theories with
matter multiplets in the fundamental representation by taking into
account non-perturbative effects. 
For asymptotically free  theories such as  pure $\CN=2$ SUSY Yang-Mills
theories the $S$-duality relation even  more involved.
The theory that is dual to microscopic non-abelian
$\CN=2$ Yang-Mills theory is an abelain theory with
monopoles. Heuristically the duality transformation in this case goes as follows.
Generically on the  moduli space of vacuums  the non-diagonal
components  of the
gauge fields are massive  and can be ``integrated out'' in the
effective low-energy description. However 
near points of  the moduli space where 
non-perturbative monopole solutions become massless 
(and  the proper description is in terms
of the dual abelain gauge fields) to obtain non-singular 
description one shall ``integrate in'' the near-massless monopole field.
The resulting theory of  dual abelian gauge fields interacting with  
monopole hypermultiplets can  be considered as an $S$-dual description
of the original pure $\CN=2$ Yang-Mills
theory. The case of finite $\CN=4$ Yang-Mills theory can be also treated
this way by  ``integrating in'' the non-diagonal gauge fields
for the dual gauge group. 

This description of $S$-duality in
four dimensions is completely analogous to the description of
the mirror symmetry for $\IP^{\ell}$-sigma models  
realized as $U(1)$-gauged linear sigma models of the fields $(X^i,\Sigma_a)$
where $X^i$, $i=1,\ldots, (\ell+1)$  are chiral superfields 
and $\Sigma$ is a twisted chiral superfield.
By integrating out  $X^i$  one obtains an  effective
theory of twisted chiral superfield $\Sigma$ with the superpotential
of the form 
\be
W(\Sigma)=\Sigma\ln \Sigma+...
\ee
Now we can ``integrate in'' additional  twisted chiral supermultiplets  $Y^j$ 
to obtain the dual  effective twisted potential 
\be
W(\Sigma,Y)=\Sigma(\sum_jY^j-r^2)+\sum_{j=1}e^{Y_j},
\ee
(see \cite{AV} for details). This provides a mirror dual
Landau-Ginzburg description of $\CN=2$ SUSY $\IP^{\ell}$ sigma model. 

We are interested in calculation of equivaraint symplectic volume of
instantons  on non-compact four-manifold \eqref{eqvol4}. Let us
outline the corresponding dual description of the SUSY gauge theories on
non-compact manifolds.  The  Seiberg-Witten solution \cite{SW}
of the  pure  $\CN=2$ SUSY
$SU(N+1)$ gauge field theory
specifies  a prepotential $\CF(\CA)$  of the low-energy effective theory 
depending on $\CN=2$ abelian vector
superfields $\CA^i$,  $i=1,\ldots, N+1$. 
The classical contribution to the prepotential $\CF(\CA)$ is given by
$\CF_0(A)=\frac{1}{2}\tau_0\sum_{i=1}^NA_i^2$, 
$\tau_0=\frac{\theta}{2\pi}+\imath \frac{4\pi}{g^2}$  and 
the complete  function $\CF(A)$ entering the Seiberg-Witten solution encodes a 
geometry of a family of algebraic curves. The corresponding action
functional can be written  in terms of $\CN=1$ chiral and vector superfields $A^i$ and
$W^i_\alpha$ as follows:
\be\label{N2A}
S=\frac{1}{4\pi}{\rm Im}\left[\int\, d^4x\,d^4\theta \,\frac{\pr \CF(A)}{\pr
    A^i}\,\bar{A}^i+\frac{1}{2}
\int \,d^4x\,d^2\theta \,\frac{\pr^2\CF(A)}{\pr A^i\pr A^j}
W^i_{\a}W^j_{\a}\right].
\ee
 Note that 
the action \eqref{N2A}  the  $\CN=2$  vector
multiplet with a general prepotential $\CF(\CA)$ can be considered as
an integral of a four-observable $\CO^{(4)}$ 
constructed from a zero-observable $\CO^{(0)}=\CF(\phi)$
by the standard descent procedure $d\CO^{(n)}=\delta \CO^{(n+1)}$.
Similar to considerations in \cite{GLO3}.    
On the non-compact four-manifold there is a non-trivial boundary
contribution breaking  $\CN=2$ SUSY invariance of the theory (this is an
analog of the Warner problem in two-dimensional SUSY theories)
\be
\delta S=\int_M d\CO^{(3)}=\int_{N=\pr M}\left(
\frac{\pr^2 \CF}{\pr \phi^i\pr \phi^j}F^i \psi^j
+\frac{1}{3!}\frac{\pr^3\CF}{\pr\phi^i\pr\phi^j\pr\phi^k}\psi^i\psi^j\psi^k
\right),
\ee
where $\phi^i$ and $\psi^i$ are components of the abelian
$\CN=2$ supermultiplet $\CA^i$
consisting of  abelain vector field 
$A^i_{\mu}$, two Weyl fermions $\lambda^i$ , $\psi^i$ and a complex scalar
field $\phi^i$. We also denote $F^i$ the curvature of the gauge field
$A^i$.  Consider now the case of $M=D\times \Sigma$, $N=\pr
M=S^1\times \Sigma$ with an action of $S^1$ by
rotations of the disk $D$.  
In $S^1$-equivariant case, 
similarly to the  two-dimensional case  \cite{GLO3}, \cite{GLO4}, 
 this boundary contribution can be canceled 
 by the  variation of a boundary term expressed through the
 Seiberg-Witten prepotential $\CF(\phi)$. In two dimensions this
boundary term enters  the  expression of the integrand of the Givental type
finite dimensional integral representation of the correlation function in topological
sigma models. 
In the case of gauge fields in four-dimensional  one expects that the Seiberg-Witten
prepotential  provides an effective description of the
integrand of the corresponding integral representation of the
equivariant symplectic volume \eqref{eqvol4}. It is reasonable to
guess that the precise description of the
integrand in the four-dimensional analog of the Mellin-Barnes integral
representation  should be given in terms of the dual theory of monopoles
interacting with the dual abelian gauge fields.

\subsection{Integral representations of vortex  counting functions}

In type $B$ topological Landau-Ginzburg sigma-models
correlation functions are naturally given by periods of
holomorphic differential forms. Thus for instance the dual type $B$ description 
of the  type $A$ topological sigma model on a disk with the target space $\IP^{\ell}$ 
leads to a finite-dimensional integral representation of the corresponding Whittaker
function \cite{GLO4}. The integrand of this integral representation is directly
related with the superpotential of the dual Landau-Ginzburg theory.
Note that the arising Whittaker function  is closely related to 
the instanton counting functions in the corresponding $\IP^{\ell}$ 
two-dimensional sigma model \cite{Gi1} (see also \cite{GLO1}). 

Taking into account the analogy between counting instantons in two and
four dimensions 
one might  expect that the instanton counting
function of \cite{LNS} can be also recasted in the compact integral form to provide a
direct relation with the Seiberg-Witten prepotential as it was
outlined at the end of the previous Subsection 
(for a direct comparison of the asymptotic of the instanton
counting function with Seiberg-Witten solution see \cite{N}, \cite{NO}). 
We postpone the construction of this integral
representation to another occasion while in the rest of this  note we
consider a degenerate version of the   instanton counting function
responsible for counting of two-dimensional vortexes (see
e.g. \cite{JT} for discussion of vortexes). 
The main result of this section is the compact Mellin-Barnes type integral
representation \eqref{intrep} of the vortex counting function \eqref{VCF}.  

Recall that the instanton counting function (up to the classical  and one loop
contributions)  in $\CN=2$ SUSY $SU(N)$-gauge theory interacts with $N_f$
hypermultiplets in the fundamental representation can be written
in the form of the infinite series \cite{LNS} 
\be
\CZ^{inst}(a,\tau,\omega,m)=1+\sum_{k=1}^{\infty}e^{2\pi \imath \tau k}\,
\CZ^{inst}_k(a,m,\omega),
\ee
where
\be
\CZ^{inst}_k(a,\omega,m)=\frac{1}{k!}\frac{(\omega_1+\omega_2)^k}
{(\omega_1\omega_2)^k}\int
\prod_{j=1}^k\frac{d\phi^j}{2\pi\imath}\,\prod_{i<j}
\frac{(\phi_i-\phi_j)^2((\phi_i-\phi_j)^2-(\omega_1+\omega_2)^2}
{((\phi_i-\phi_j)^2-\omega_1^2)((\phi_i-\phi_j)^2-\omega_2^2)}\times 
\ee
$$
\times
\,\,\prod_{j=1}^k\frac{\prod_{\a=1}^{N_f}(\phi_j+m_\a)}
{\prod_{l=1}^N(\phi_j-a_l)(\phi_j-a_l+\omega_1+\omega_2)}. 
$$
The vortex counting function (see e.g. \cite{Shad}) can be defined by taking
a limit of the instanton counting function  
\be
\CZ^{vortex}(a,\tau,m,\omega_1)=\lim_{\omega_2\to \infty}\,
\CZ^{inst}(a,\tau+\frac{N}{2\pi \imath}\ln\omega_2,m,\omega_1,\omega_2).
\ee
The limit can be taken explicitly to obtain (we use 
simplify notations $\omega:=\omega_1$ below)
\be\label{VCF1}
\CZ^{vortex}(a,\tau,\omega,m)=1+\sum_{k=1}^{\infty}e^{2\pi \imath \tau k}\,
\CZ_k(a,m,\omega),
\ee
where
\be\label{vc}
\CZ_k(a,\omega,m)=\frac{1}{k!}\frac{1}{\omega^k}\int
\prod_{j=1}^k\frac{d\phi^j}{2\pi\imath}\,\prod_{i< j}^k\frac{(\phi_i-\phi_j)^2}
{(\phi_i-\phi_j)^2-\omega^2}\,\,\prod_{j=1}^k\frac{\prod_{\a=1}^{N_f}(\phi_i+m_\a)}
{\prod_{l=1}^N(\phi_j-a_l)}.
\ee
Here the integration goes over $\IR^k$ and we imply that $a_l$ and
$\omega$ have small  positive imaginary parts. Let us multiply
\eqref{VCF1}  by a perturbative contribution $\CZ^{pert}$ 
\be\label{VCF}
\CZ(a,\tau,m,\omega)=\CZ^{pert}(a,\tau,m,\omega)\CZ^{vortex}(a,\tau,m,\omega),
\ee
\be \label{VCF2}
\CZ^{pert}(a,\tau,m,\omega)
=\frac{\prod_{\a=1}^{N_f}\prod_{p=1}^N\Gamma_1(a_p+m_\a+\omega|\omega)}
{\prod_{p\neq q}^N\Gamma_1(a_p-a_q|\omega)}.
\ee
The integral \eqref{vc} over $k$-vortex moduli space
can be expressed as a sum over residues 
(see \cite{N}, \cite{NO} for similar calculations  in the case of instantons)
\be\label{sumover}
\CZ_k(a,\omega)=\sum_{|\underline{k}|=k}\frac{1}{\underline{k}!\omega^k}
\frac{\prod_{f=1}^{N_f}\prod_{p=1}^N\prod_{i_p=1}^{k_p}(a_p+m_f+(i_p-1)\omega)}
{\prod_{l\neq m}^N\prod_{i_l=1}^{k_l}(a_l-a_m+(k_l-k_m-i_l)\omega)},
\ee
where the sum goes over partitions $\underline{k}=(k_1,k_2,\ldots ,k_{N})$,  $k_l\in
\IZ_{\geq 0}$, $|\underline{k}|=k_1+k_2+\ldots +k_N$. 
Taking into account  the following simple identities
\be
\frac{\Gamma(x+n+1)}{\Gamma(x)}=\prod_{p=0}^n(x+p),\qquad
\frac{\Gamma(x)}{\Gamma(x-n)}=\prod_{p=1}^{n}(x-p),\quad n>0,
\ee
\be
\frac{\pr}{\pr x}\left(\frac{1}{\Gamma(x-k)}\right)|_{x=0}=(-1)^k\,k!\,,
\ee
 we have
\be
\frac{1}{\prod_{l\neq
  m}^N\prod_{i_l=1}^{k_l}(a_l+(k_l-i_l)\omega-(a_m+k_m\omega))}=
\frac{\prod_{l\neq m}^N\Gamma_1(a_l-(a_m+k_m\omega)|\omega)}
{\prod_{l\neq  m}^N \Gamma_1(a_l+k_l\omega-(a_m+k_m\omega)|\omega)},
\ee
\be
\prod_{f=1}^{N_f}\prod_{p=1}^N\prod_{i_p=1}^{k_p}(a_p+m_f+(i_p-1)\omega)=
\prod_{f=1}^{N_f}\prod_{p=1}^N\frac{\Gamma_1(a_p+m_f+k_p\omega|\omega)}
{\Gamma_1(a_p+m_f+\omega|\omega)}.
\ee
Thus  the vortex generating function can be rewritten as follows:
\be\label{INF}
\CZ^{vortex}(a,\tau,m,\omega)=\frac{1}{\prod_{f=1}^{N_f}\prod_{p=1}^N
\Gamma_1(a_p+m_f+\omega|\omega)}
\sum_{k=0}^{\infty}e^{2\pi \imath \tau
  k}\sum_{|\underline{k}|=k}\frac{1}{\underline{k}!\omega^k}
\ee
$$
\times \prod_{f=1}^{N_f}\prod_{p=1}^N\Gamma_1(a_p+k_p\omega+m_f|\omega)
\prod_{l\neq m}^N\frac{\Gamma_1(a_l-(a_m+k_m\omega)|\omega)}
{ \Gamma_1(a_l+k_l\omega-(a_m+k_m\omega)|\omega)}.
$$
Let us consider  the following multiple integral 
\be\label{intrep}
\widetilde{\CZ}(a,\tau,m,\omega)=
\int_{\CS} \prod_{j=1}^N\frac{d\phi_j}{2\pi\imath}\,  e^{2\pi
    \imath (\tau+\frac{1}{2})\sum_{j=1}^{N}(\phi_j-a_j)/\omega}\,\,\,
\frac{\prod_{j=1}^N\prod_{l=1}^N
  \Gamma_1(a_l-\phi_j|\omega)}
 {\prod_{i\neq j}\Gamma_1(\phi_i-\phi_j|\omega)}
\times
\ee
\be\nonumber
\times \frac{\prod_{j=1}^N\prod_{\a=1}^{N_f}\Gamma_1(\phi_j+m_\a|\omega)}
 {\prod_{j\neq i}\Gamma_1(a_j-a_i|\omega)}, 
\ee
where the integration contour $\CS$ encloses only the poles of
$\Gamma_1(a_l-\phi_j|\omega)$ (i.e we take $a_l\to a_l+\i0$ and
$m_f\to m_f+\i0)$. The integral is given by as sum over 
residues at $\phi_j=a_l+k_l\omega$ such that
different $\phi_j$ correspond to different $a_l$.
Taking into account that the integrand  is symmetric with respect to
interchange of $\phi_j$ we can take,  up to a simple symmetric factor,  
$\phi_j=a_j+k_j\omega$. Then the evaluation of the residues leads
to identification of the integral expression
$\widetilde{\CZ}(a,\tau,m,\omega)$ given by \eqref{intrep} 
with the vortex counting function \eqref{VCF}, \eqref{VCF2}, \eqref{INF}.

\vskip 1cm

\noindent {\small {\bf A.G.} {\sl Institute for Theoretical and
Experimental Physics, 117259, Moscow,  Russia; \hspace{8 cm}\,
\hphantom{xxx}  \hspace{2 mm} School of Mathematics, Trinity College
Dublin, Dublin 2, Ireland; \hspace{6 cm}\hspace{5 mm}\,
\hphantom{xxx}   \hspace{2 mm} Hamilton Mathematics Institute,
Trinity College Dublin, Dublin 2, Ireland;}\\

\noindent{\small {\bf D.L.} {\sl
 Institute for Theoretical and Experimental Physics,
117259, Moscow, Russia};\\

\end{document}